\documentclass[10pt,pra,twocolumn,floatfix,showpacs%,superscriptaddress
]{revtex4-1}
\usepackage{amsfonts}
\usepackage{amssymb}
\usepackage{amsmath}
\usepackage{mathbbol}
\usepackage{graphicx}
\usepackage{setspace}
\usepackage{xcolor}

\def\eq#1{Eq.~(\ref{eq:#1})}

\begin{document}

\title{Spatiotemporal graph states from a single optical parametric oscillator}

\author{Rongguo Yang}
\author{Jing Zhang}
\email{zjj@sxu.edu.cn}
\affiliation{College of Physics and Electronic Engineering, Collaborative Innovation Center of Extreme Optics, Shanxi University, Taiyuan 030006, P.R. China}
\affiliation{Department of Physics, University of Virginia, 382 McCormick Road, Charlottesville, Virginia 22904-4714, USA}
\author{Israel Klich}
\author{Carlos Gonz\'alez-Arciniegas}
\author{Olivier Pfister}
\email{olivier.pfister@gmail.com}

\affiliation{Department of Physics, University of Virginia, 382 McCormick Road, Charlottesville, Virginia 22904-4714, USA}

\date{\today}

\begin{abstract}
 An experimental scheme is proposed for building massively multipartite entangled states using both the spatial and the frequency modes of an optical parametric oscillator. 
We provide analytical forms of the entangled states using the squeezed eigenmodes of Heisenberg equations, a.k.a.\ the nullifiers of the corresponding graph state. 
This scheme can generate,  in parallel, several cluster states described by sparsely connected, bicolorable  graph states, usable for one-way quantum computing. 
We indicate the experimentally accessible quantum graphs, depending on the squeezing parameter.
\end{abstract}

\maketitle

\section{Introduction}
Quantum entanglement is an elementary and key resource for quantum information and quantum computing~\cite{Nielsen2000}. Cluster states, as 2D sparse graph states, are of central importance for one-way quantum computing, be it over discrete variables~\cite{Briegel2001,Raussendorf2001} or continuous variables (CV)~\cite{Zhang2006,Menicucci2006,Gu2009}. It is important to note that CV quantum computing~\cite{Pfister2019} is a valid type of universal quantum computing that features the same exponential speedup~\cite{Lloyd1999,Bartlett2002} as well as a quantum error correction encoding~\cite{Gottesman2001} and a fault tolerance threshold~\cite{Menicucci2014ft}.
 
The experimental generation of photonic CV cluster states first used ``bottom up'' quantum-circuit like approaches~\cite{vanLoock2007}, based on the Bloch-Messiah decomposition~\cite{Braunstein2005} which yielded  four-mode~\cite{Su2007,Yukawa2008a} and eight-mode~\cite{Su2012} cluster states, using several optical parametric oscillators (OPOs) and  linear optical transformations. In this approach the number of OPOs is proportional to the number of entangled modes. 

An alternative, “top down” approach was proposed and realized, first in the frequency domain~\cite{Pfister2004,Menicucci2007} then in the time domain~\cite{Menicucci2010,Menicucci2011}. Such an approach requires only one OPO, or two, to generate two-mode squeezed states, a.k.a.\ Einstein-Podolsky-Rosen (EPR) pairs, over the OPO's optical frequency comb or, alternatively, in temporally pulsed modes. The first proposal featured a square-grid cluster state, universal for quantum computing, but required a single OPO with a triple (but demonstrated~\cite{Pysher2010}) nonlinear medium and, in particular, a complex pump spectrum~\cite{Menicucci2008}. Subsequent, simpler proposals for building cluster states sequentially in the time domain~\cite{Menicucci2010,Menicucci2011} were adapted experimentally in the frequency domain to build 15 independent quadripartite square cluster states~\cite{Pysher2011} and one 60-partite, and two 30-partite, one-dimensional cluster states~\cite{Chen2014}. Note that, in the latter case, the demonstrated number of entangled qumodes was not limited by the OPO phasematching bandwidth, which extends to at least 6700 modes~\cite{Wang2014}, but by the tunability of the local oscillator laser in the interferometric squeezing measurements. Another approach used a synchronously pumped OPO to yield a much broader qumode frequency spacing, well suited for parallel quantum processing~\cite{Roslund2014}. In the temporal domain, entangled modes are obtained sequentially, only two or four at a time but the resulting polynomial overhead for quantum processing is offset by scalability, the total number of entangled modes being limited only by the stability of the experiment. One-dimensional cluster states were thus generated over $10^{4}$~\cite{Yokoyama2013}, then $10^{6}$~\cite{Yoshikawa2016} modes. Recently, large-scale two-dimensional square-lattice cluster states were generated~\cite{Asavanant2019,Larsen2019}. 

In this context, an interesting degree of freedom to explore, and add,  is the transverse spatial one. Continuous-variable entanglement between two spatial modes was realized within one beam~\cite{Janousek2009}. Linear cluster states were produced among different spatial modes and all possible spatial modes of light were copropagated within one beam~\cite{Armstrong2012}. The generation of a CV dual-rail cluster state based on an optical spatial mode comb was proposed via a four-wave-mixing process~\cite{Pooser2014}. Proposals were also made for large-scale CV dual-rail cluster state generation involving Laguerre-Gaussian (LG) modes in a large-Fresnel-number degenerate OPO~\cite{Yang2016}, and in a spatial mode comb pumped by two spatial LG modes~\cite{Zhang2017}. A CV square quadripartite  cluster state was experimentally produced by multiplexing orthogonal spatial modes in a single optical parametric amplifier (OPA) ~\cite{Cai2018}.
\begin{figure*}[htb]
\begin{centering}
\includegraphics[width=.8\textwidth]{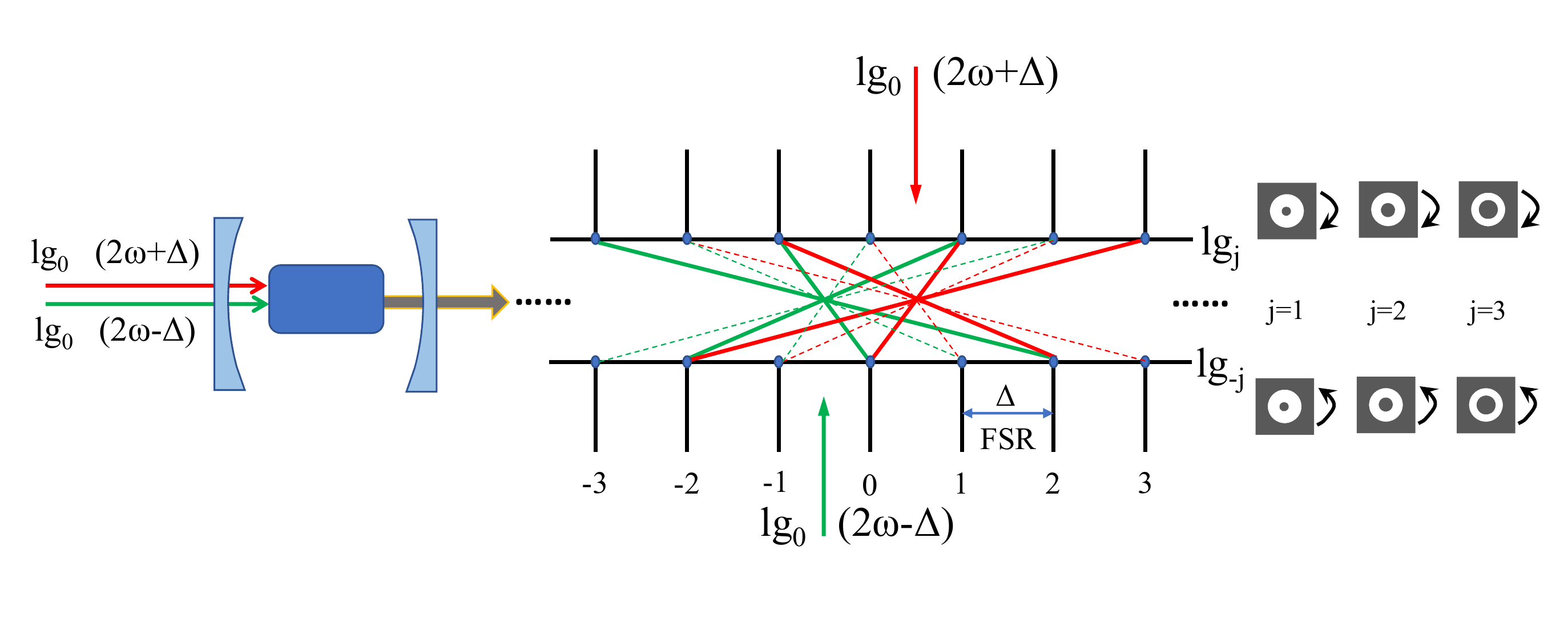}
\end{centering}
\caption{Schematic of experimental setup. the green and red arrows represent pump modes ${\lg _0}$ with the frequency ${\omega _p} = 2\omega  \pm \Delta $, the green (red) solid/dashed (generate symmetrically) line connect two modes of down-converted process from the same pump.}
\label{fig1}
\end{figure*}

In this paper, we consider the process of optical parametric amplification in a single OPO pumped by two LG modes and the parallel generation of entangled graph states in both the frequency and space domains. This paper is organized as follows: In Sec.~II, we describe our system and solve its Hamiltonian analytically to calculate the resulting multipartite entanglement using the CV graph state formalism through the $G$ and $A$ adjacency matrices for the $\mathcal H$ graph and canonical graph, respectively~\cite{Menicucci2007}. In Sec.~III, we give numerical illustrations of this result for 8 and 60 modes, and indicate how finite squeezing would affect measurements of weighted-graph states. 

\section{Physical system and analytic solutions}
In our system, a nonlinear crystal is placed in a self-imaging, or large Fresnel number, cavity, whose transverse eigenmodes are the complete set of LG modes, and longitudinal eigenmodes are spaced by the free spectral range (FSR). This specially designed cavity can guarantee simultaneous and sustainable nonlinear interaction and resonance of all the down converted modes~\cite{NavarreteBenlloch2009}.  As is shown in Fig.1, the system is pumped by two spatial LG modes ${\lg _0}$ with frequencies ${\omega _p} = 2\omega  \pm \Delta $, where $\Delta$ is the FSR, and $\mu=\pm j$ in ${\lg _\mu}$  mode denotes the OAM mode number. Here $j$= 0 for the pump field and $j$ =   1,  2,  3, ..., m for the corresponding downconverted fields. The nonlinear crystal in the cavity is a type I phase-matching crystal whose second order nonlinear coefficient is $\xi$. The two pump fields of frequency $\omega_p$ can be downconverted into signal and idler fields of frequencies $\omega _{s,i}$=$\omega  \pm n\Delta$, $m$ = 0,1,2... The nonlinear interaction must satisfy energy conservation ($\omega _p$ =  $\omega _s$ +  $\omega _i$), phase-matching  ($\vec {{k_p}}$  = $\vec {{k_s}}$  + $\vec {{k_i}}$) and orbit angular momentum conservation (${\mu_p=0}$  = ${\mu_s}$  + ${\mu_i}$). Each pair of modes connected by a red or green line are from the same optical down conversion process and can form a high-connected entanglement state of optical frequency and spatial modes.

The interaction Hamiltonian of the system is
%\begin{align}\label{eq:H}
% H=i\hbar \xi &\sum\limits_{i =  - \frac n2}^{\frac n2}\sum\limits_{j = 1}^{j}\{ {{G_{( - i, \pm j),( - 1 + i, \mp j)}}} {{\hat b}_{ - 1,0}}\hat a_{ - i, \pm j}^\dag \hat a_{ - 1 + i, \mp j}^\dag  \nonumber\\
% &+ {{G_{(i, \pm j),(1 - i, \mp j)}}} {{\hat b}_{1,0}}\hat a_{i, \pm j}^\dag \hat a_{1 - i, \mp j}^\dag\} +H.c.,  
% \end{align}
\begin{align}\label{eq:H}
 H=i\hbar \xi &\sum\limits_{i=-n}^{n}\sum\limits_{j=1}^{m}\{ {{G_{( - i, \pm j),( - 1 + i, \mp j)}}} {{\hat b}_{ - 1,0}}\hat a_{ - i, \pm j}^\dag \hat a_{ - 1 + i, \mp j}^\dag  \nonumber\\
 &+ {{G_{(i, \pm j),(1 - i, \mp j)}}} {{\hat b}_{1,0}}\hat a_{i, \pm j}^\dag \hat a_{1 - i, \mp j}^\dag\} +H.c.,  
 \end{align}
where the $G$ matrix element is 1 when the parametric process exists and 0 otherwise. ${\hat b_{1,0}}$ and ${\hat b_{-1,0}}$ denote the annihilation operators of the two pump modes with frequency $2\omega  \pm \Delta$, respectively,  their first and second  index corresponding to the frequency and OAM number, respectively. The pump ${\hat b_{-1,0}}$ can thus produce ${\hat a_{ - i, \pm j}}$ and ${\hat a_{ - 1 + i, \mp j}}$ by down-conversion process, and the pump ${\hat b_{1,0}}$ can produce ${\hat a_{i, \pm j}}$ and ${\hat a_{1 - i, \mp j}}$.  Here, we only consider the first order LG modes, $j$ = 1, $\mu$=$\pm1$, in the down-conversion modes; similar reasoning can be used when considering higher order modes $j>$1. The physical system and corresponding $\mathcal H$ graph are shown in Fig.2. In Fig. 2(a), certain upper modes of ${\lg _1}$ and certain lower modes of ${\lg _{ - 1}}$ are connected by the red (green) lines, and the cross point of all red (green) lines corresponds to the red (green) pump. In Fig. 2(b), one can see the relation and structure of the down-converted fields clearly.
\begin{figure}[htb]
\begin{centering}
\includegraphics[width=\columnwidth]{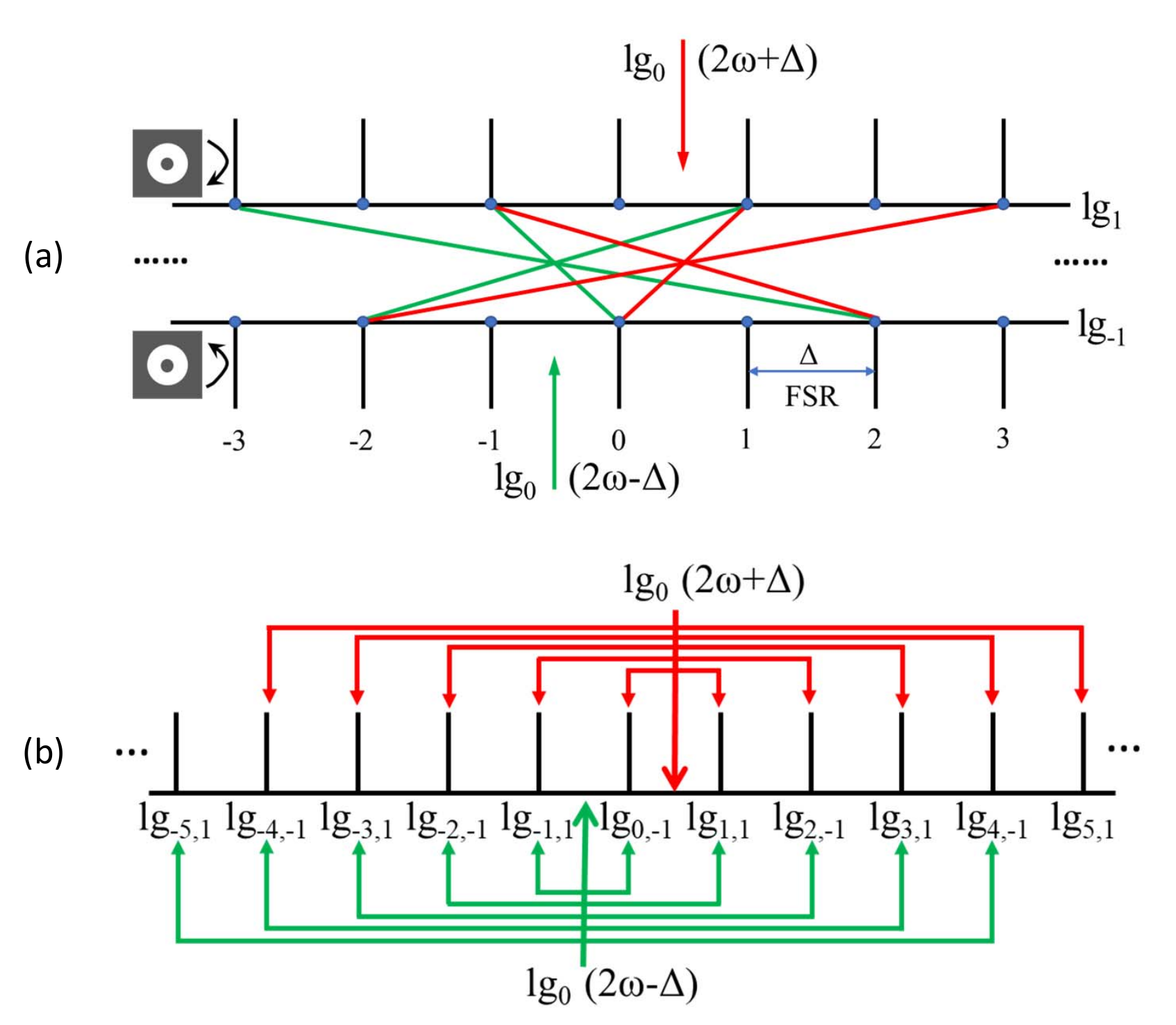}
\end{centering}
\caption{(a) Physical picture of the mode entanglement with different pumps. Downconverted modes ($j$=1 only) from each pump are connected with red  and green arrows. (b) $\mathcal H$ graph.}
\label{fig2}
\end{figure}
For convenience, we rename these modes from 1 to $N=2n$ (we always consider a even number of modes). According to the interaction Hamiltonian and the renamed modes, the G matrix can be written as
\begin{equation}
G = \left( {\begin{array}{*{20}{c}}
	{} &  \vdots  & {}  \\
	\cdots  & {\begin{array}{*{20}{c}}
		
		0 & 0 & 0 & 0 & 0 & 0 & 1 & 0  \\
		0 & 0 & 0 & 0 & 0 & 1 & 0 & 1  \\
		0 & 0 & 0 & 0 & 1 & 0 & 1 & 0  \\
		0 & 0 & 0 & 0 & 1 & 1 & 0 & 0  \\
		0 & 0 & 1 & 1 & 0 & 0 & 0 & 0  \\
		0 & 1 & 0 & 1 & 0 & 0 & 0 & 0  \\
		1 & 0 & 1 & 0 & 0 & 0 & 0 & 0  \\
		0 & 1 & 0 & 0 & 0 & 0 & 0 & 0  \\
		\end{array}} &  \cdots   \\
	{} &  \vdots  & {}  \\
	\end{array}} \right).
\label{eq:G}
\end{equation}
for any $N$-mode state. This matrix can be seen as the adjacency matrix of the $\mathcal H$ graph, which is bicolorable. The spectrum of this graph can be derived analytically (see Appendix for details). The eigenvalues of $G$ are 
\begin{align}
\lambda_k &= \pm 2\left|\cos\frac{k\pi}{2n+1}\right|,
\end{align}
where  $k=1,2,3,...,n$. Because there are $n$ eigenvalues with each sign,  the solutions of the Heisenberg equations are $n$ quadrature-amplitude squeezed  and $n$ phase-quadrature squeezed eigenmodes, of  squeezing factor $\exp(\lambda_{k}\xi t)$, $t$ being the Hamiltonian interaction time (or cavity lifetime in this simplified model).  

For pure two-mode squeezing Hamiltonians such as that of  \eq H, the multipartite state resulting from the quantum evolution can always be expressed as a cluster state~\cite{Menicucci2007}. This means that the $N$ eigenmodes always have the following quantum standard deviation
\begin{equation}
\Delta[\vec P(t)-A\vec Q(t)] \propto e^{-\xi t},
\end{equation}
where $\vec P(t)=[P_{1}(t),\dots,P_{N}(t)]^{T}$, ~$\vec Q(t)=[Q_{1}(t),\dots,Q_{N}(t)]^{T}$, $P_{j}=(a_{j}-a_{j}^{\dag})/(i\sqrt2)$, $Q_{j}=(a_{j}+a_{j}^{\dag})/\sqrt2$, and $A$ is the  adjacency matrix of the canonical cluster graph. In order to determine the  corresponding cluster state, we need to derive this adjacency matrix. This can be done analytically (see Appendix for details), yielding
\begin{equation}\label{eq:A}
A = \left( {\begin{array}{*{20}{c}}
	0 & S^TB^TSJ \\
	JS^TBS & 0 \\
\end{array}} \right),
\end{equation}
 where $J$ is the anti-diagonal identity matrix, $S$ is a permutation matrix, and $B$ is constructed from the eigenvectors of $G$.  The matrices $B$ in \eqref{eq:A} are derived explicitly in the appendix, with the result:
 \begin{align}
B_{ij}=\frac{(-1)^{i+j+n}}{1+2 n}  \left[\frac{1}{\cos (\frac{(i-j) \pi }{1+2 n})}+\frac{1}{\cos (\frac{(i+j-1) \pi }{1+2 n})}\right].
\end{align}
The matrices $S$ and $J$ have the matrix elements $S_{ij}=\delta _{j,2i-1}+\delta _{j,2i-2n-2}$ and $J_{ij}=\delta_{i,n-j+1}$.
 
The form \eqref{eq:A} of $A$ shows that the resulting cluster graph is bicolorable.

\section{Illustrative examples}

In this section, we give numerical examples of the obtained bicolorable graph states for two different scales. This assumes that the phasematching bandwidth of the OPO has a flat-top shape and ``turns off'' sharply between the last mode in the considered set and its neighbor. In practice, this is not doable and the coupling to wing modes will be tapering. However, such boundary imperfections can be considered confined to their location if the graph is sparse, i.e., local, enough. (In the case of GHZ states, which are complete graphs, this can be more of a problem.)

\subsection{Small scale}
 The possible $\mathcal H$ graphs for an eight-mode state are shown in Fig.3, with each vertex corresponding to a different qumode. Red and green lines correspond to different pumps. 
 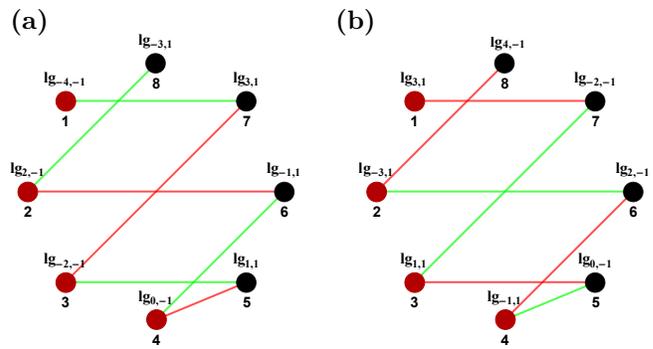
\begin{figure}[htb]
\begin{centering}
\vglue .1in
\setlength{\unitlength}{0.1\columnwidth}
\begin{picture}(10,0)
\put(0,0){\bf(a)}
\put(5,0){\bf(b)}
\end{picture}
\includegraphics[width=0.45\columnwidth]{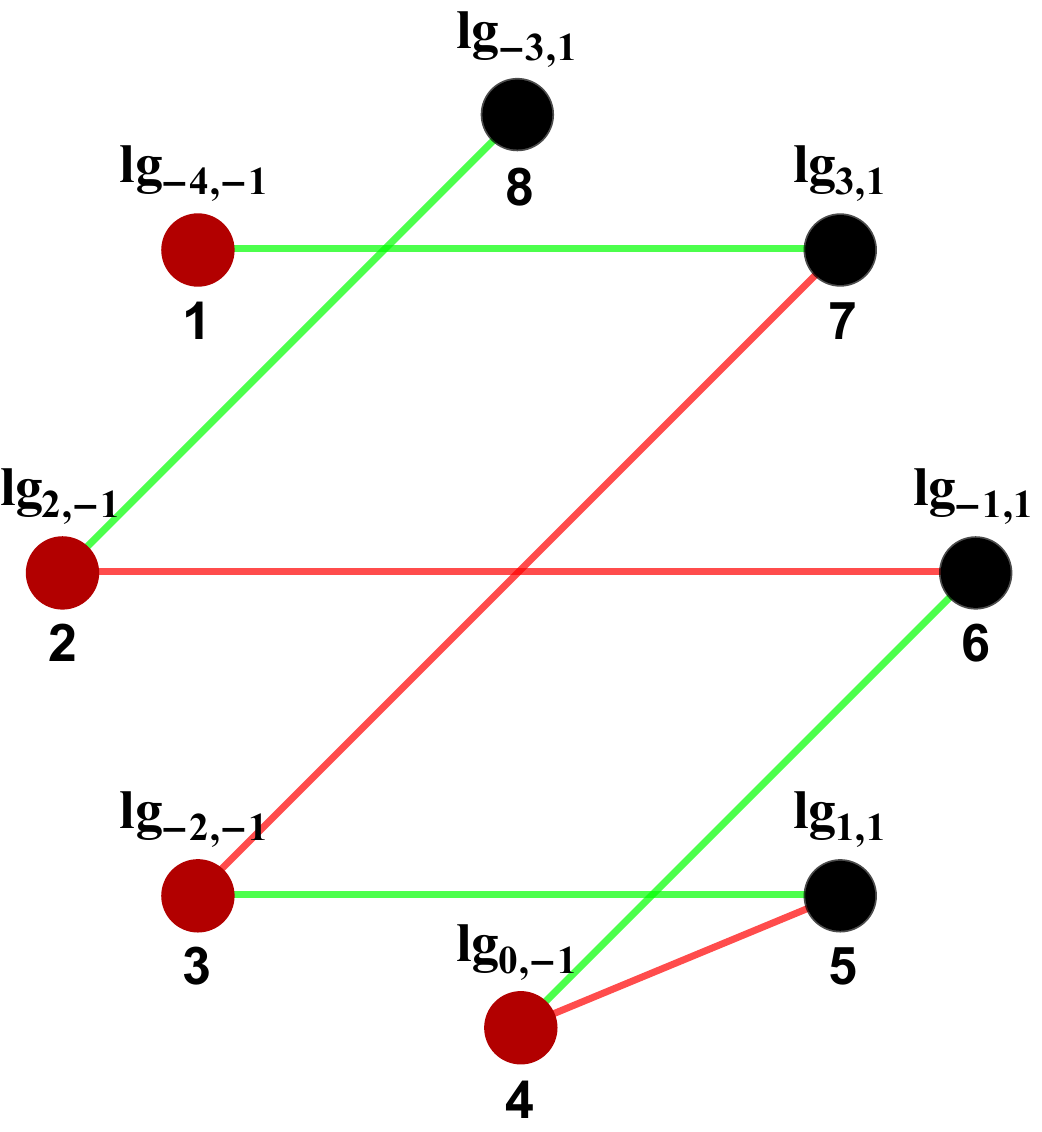}\hfill\includegraphics[width=0.45\columnwidth]{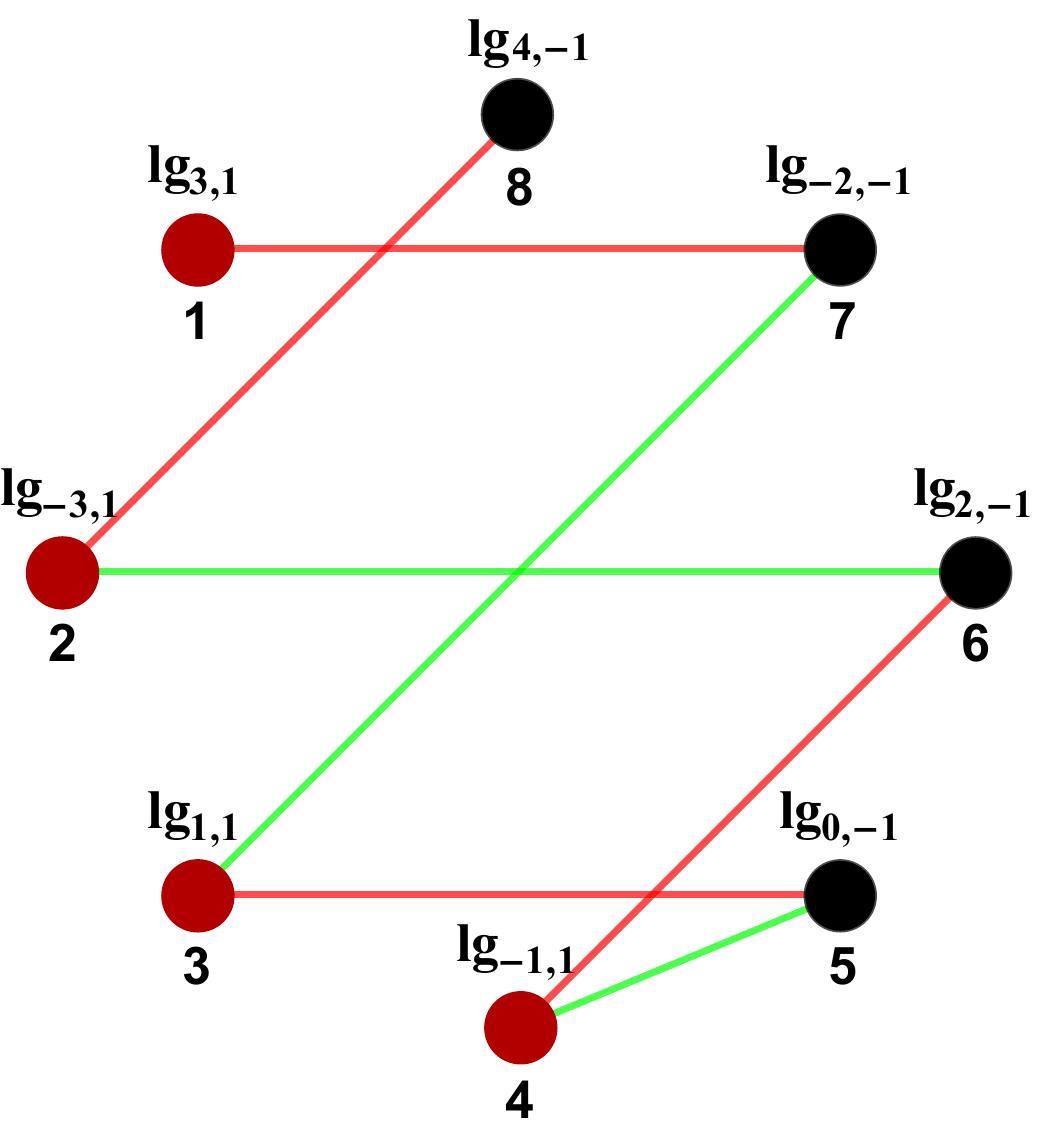}
\end{centering}
\caption{The two possible $\mathcal H$ graphs for an eight-mode state.}
\label{fig3}
\end{figure}
It is interesting to note that both these graphs have the same linear chain structure, i.e., that $G$ is a Hankel-like matrix. As shown in the previous section, the canonical (e.g., cluster) graph state is obtained from its adjacency matrix $A$, calculated from \eq A and also using the method outlined in Ref.~\citenum{Menicucci2007}, and displayed in Fig.4(a). This matrix corresponds, in general to a weighted complete bicolorable graph, drawn in Fig.4(b). 
\begin{figure}[htb]
\begin{centering}
\vglue .1in
\setlength{\unitlength}{0.1\columnwidth}
\begin{picture}(10,0)
\put(0,0){\bf(a)}
\put(5,0){\bf(b)}
\end{picture}
\includegraphics[width=0.45\columnwidth]{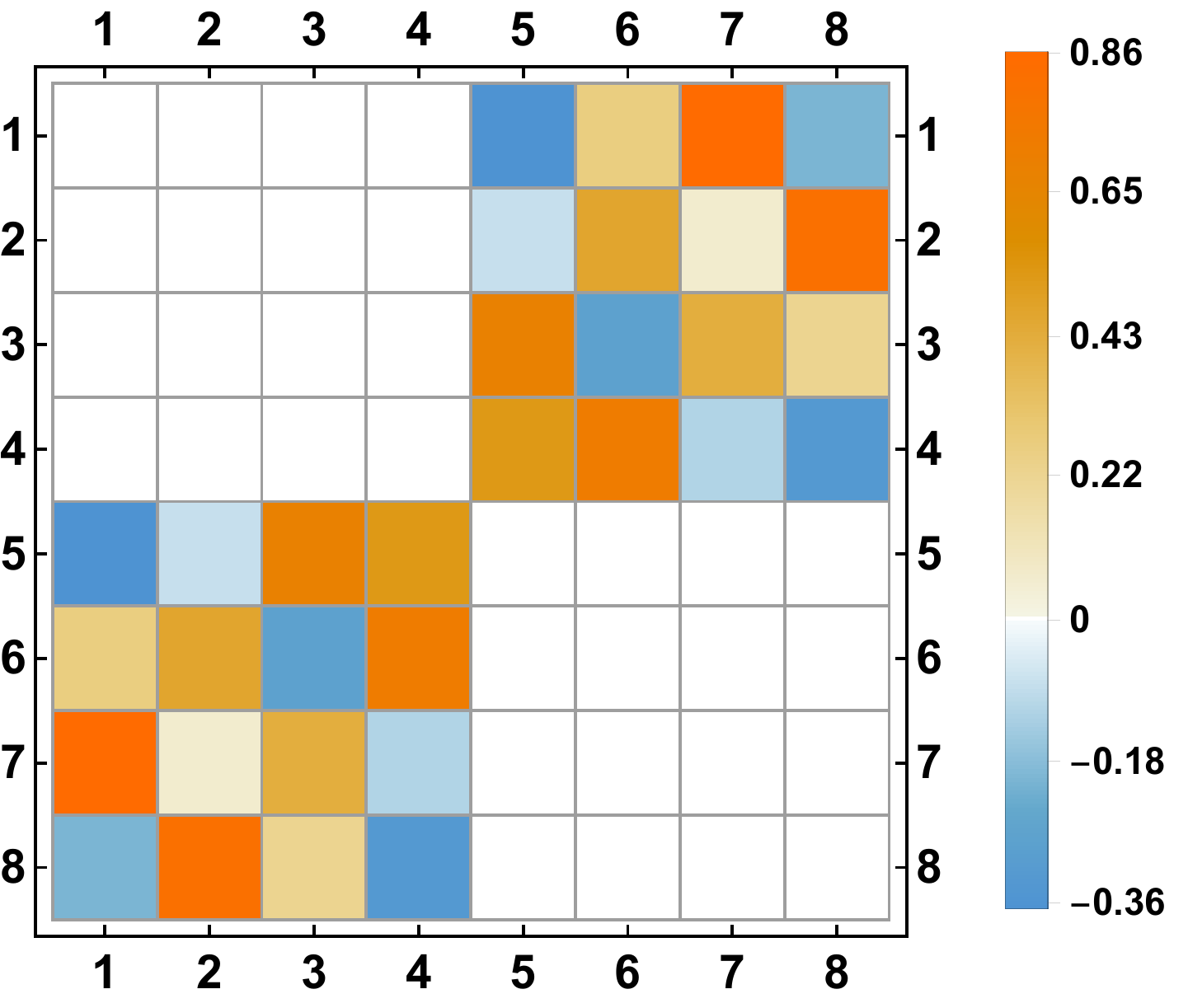}\hfill\includegraphics[width=0.45\columnwidth]{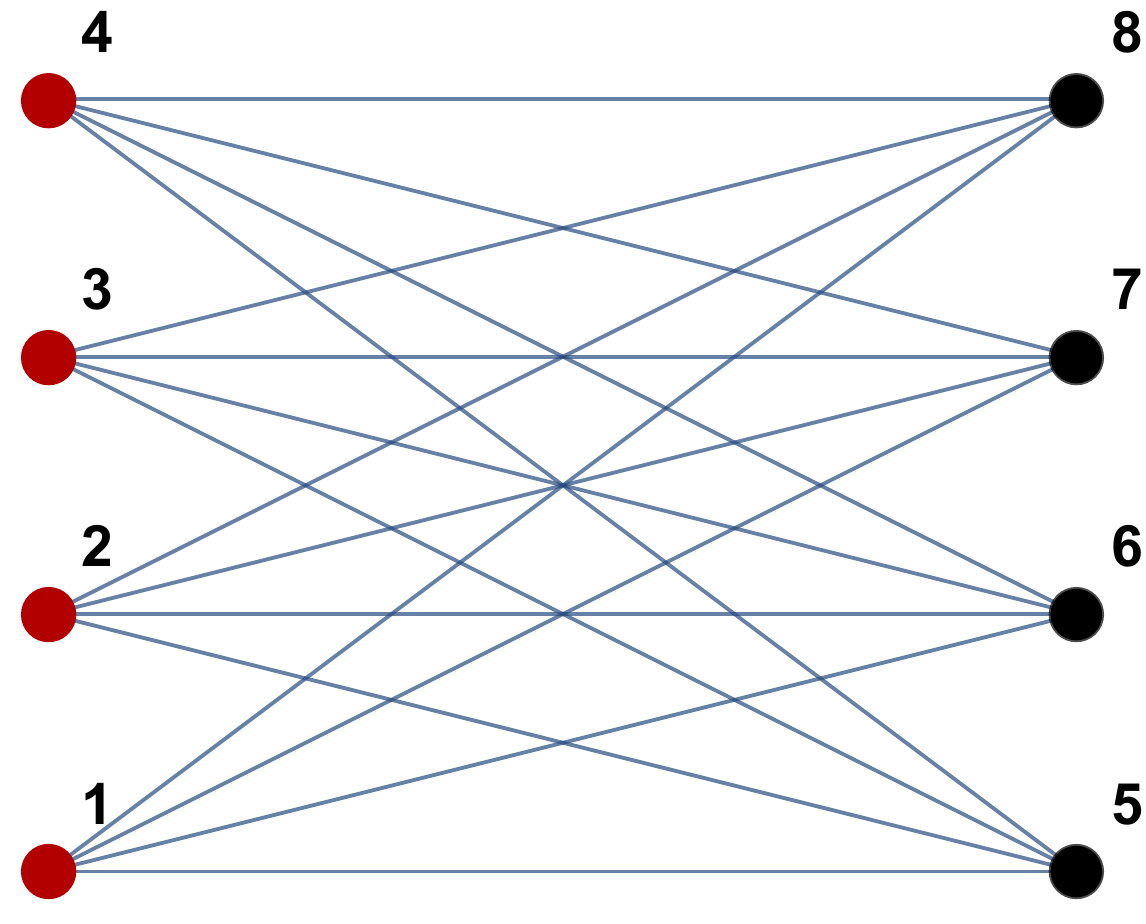}
\end{centering}
\caption{ (a),  $A$ matrix and, (b), corresponding canonical graph for an 8-mode state.}
\label{fig4}
\end{figure}
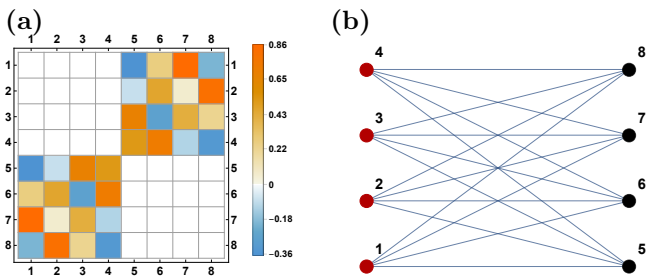

\subsection{Large scale}
In order to gain perspective on the overall structure---if any---of the graph state, we now expand to an arbitrary larger, yet computable, number of modes, e.g.\  60 modes. Note that the maximum mode number depends on the ratio of the phasematching spectral bandwidth with the FSR. The phasematching bandwidth in PPKTP can be very large, on the order of 10 THz for some interactions with a 532 nm pump wavelength, as was calculated and measured  in Ref.~\citenum{Wang2014}. For a typical FSR on the order of 1-10 GHz, this can yield $10^{3}-10^{4}$ qumodes. Figure 5 displays the $A$ matrix for 60 modes. 
\begin{figure}[hb]
\begin{centering}
\includegraphics[width=\linewidth]{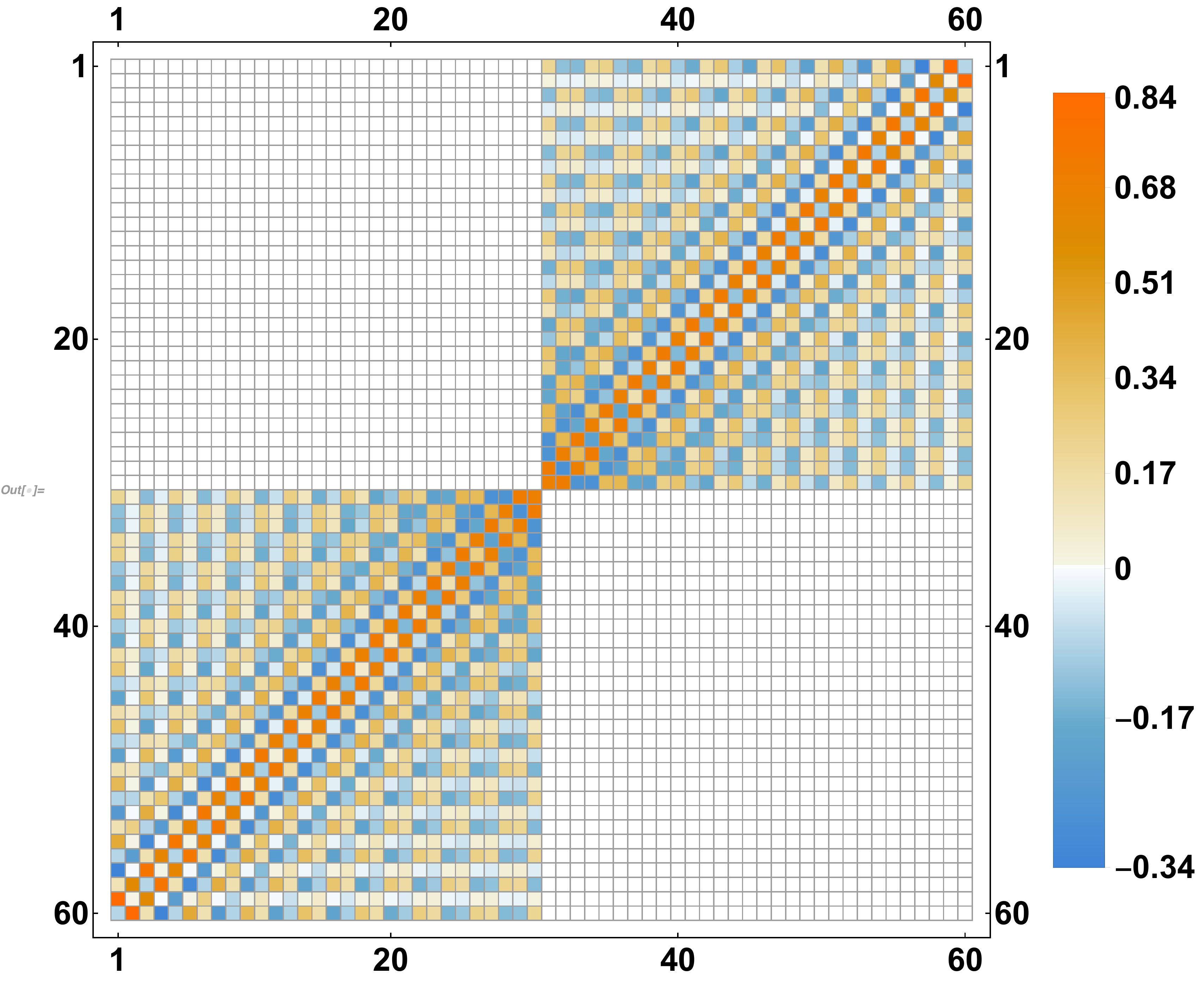}
\caption{The A matrix for 60 modes.}
\label{fig5}
\end{centering}
\end{figure}
Besides its aforementioned bipartite structure, $A$ also has a  skew-symmetric structure that clearly mirrors that of $G$, \eq G \footnote{Note that it is well known~\cite{Zaidi2008} that $A=G$ if $G^{2}=\mathbb1$, but that isn't the case here.}. Although $A$ is nominally a complete bicolorable graph (i.e., nodes 1 to 30 are not linked to one another but are all linked to all of the nodes 31 to 60), it is also strongly weighted, as the absolute values of the nonzero A matrix elements range from orders $10^{-4}$ to 1. A natural question is then to ask what is the physical significance, and even the relevance, of these very weak edges. An intuitive answer to this question is to examine the edge weight relative to the available squeezing~\cite{Xuan}: in a nutshell, it is well known that the edge between two qumodes describes their quantum correlations, as evidenced by the formal equivalence between a two-mode cluster state and a two-mode squeezed state~\cite{Menicucci2011}. The squeezing parameter $\xi t$ used in experimentally realizing cluster entanglement then naturally serves as the noise floor for observing quantum correlations  between modes and a rule of thumb is that an edge of weight $\varepsilon$, i.e., an element of $A$ of value $\varepsilon$, will only be relevant if the squeezing is large enough, i.e., if $\varepsilon\gtrsim\exp(-2\xi t)$; otherwise the quantum correlations due to the edge of weight $\varepsilon$ will be buried in the squeezed quantum noise, therefore unobservable and, for all intents and purposes, nonexistent. 

With this criterion in mind, we examine matrix $A$ again after rounding down all elements below a certain threshold to zero, the thresholds being chosen to correspond to realistic values of squeezing: Fig.6 displays the results of such ``graph pruning'' for three different, contiguous threshold ranges. Because the graphs are regular, save for local imperfections at the boundaries (chains' ends), we only displayed central sections to clearly highlight changes in graph valence and structure.
\begin{figure}[htb]
\setlength{\unitlength}{0.1\columnwidth}
\vglue 0.1in
\begin{picture}(10,0)
\put(0,0){\large\bf(a) \underline{-2.6 dB to -6.4 dB}}
\end{picture}
\centerline{\includegraphics[width=.9\columnwidth]{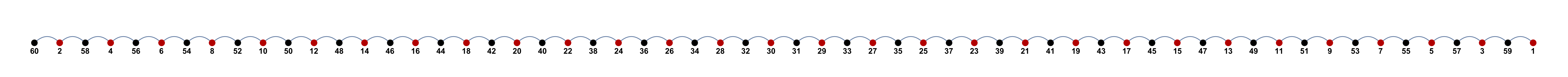}}
\begin{picture}(10,1)
\put(0,-0){\large\bf(b) \underline{-6.6 dB to -7.2 dB}}
\end{picture}
\centerline{\includegraphics[width=.9\columnwidth]{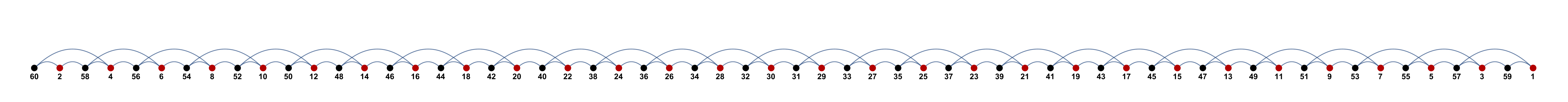}}
\centerline{\includegraphics[width=.9\columnwidth]{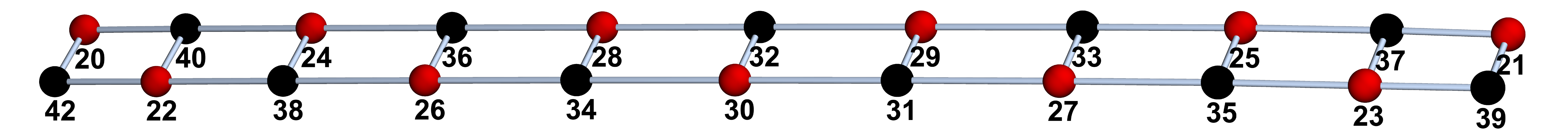}}
\begin{picture}(10,1)
\put(0,0.25){\large\bf(c)\underline{ -7.4 dB to -8.2 dB}}
\end{picture}
\centerline{\includegraphics[width=.9\columnwidth]{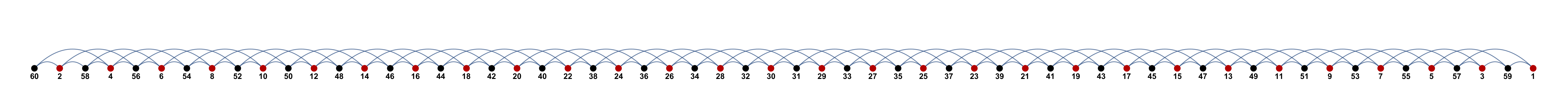}}
\centerline{\includegraphics[width=.9\columnwidth]{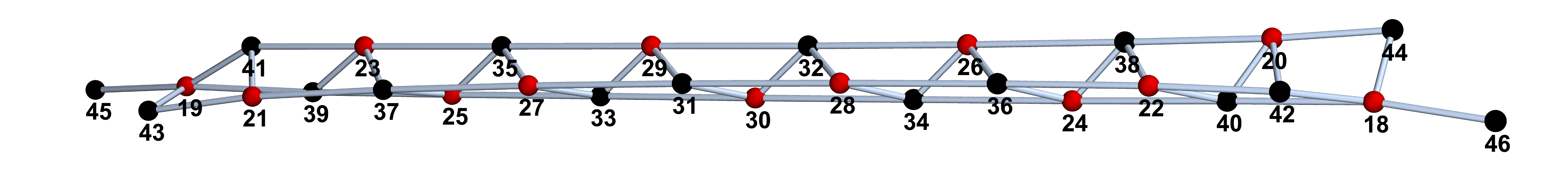}}
\caption{Graph states obtained from $A$ with increasing squeezing, by neglecting $A$ elements below threshold. (a), threshold range 0.55-0.23 (squeezing range -2.6 dB to -6.4 dB); (b) threshold range 0.22-0.19 (squeezing range -6.6 dB to -7.2 dB); (c) threshold range 0.18-0.15 (squeezing range -7.4 dB to -8.2 dB).} 
\label{fig6}
\end{figure}
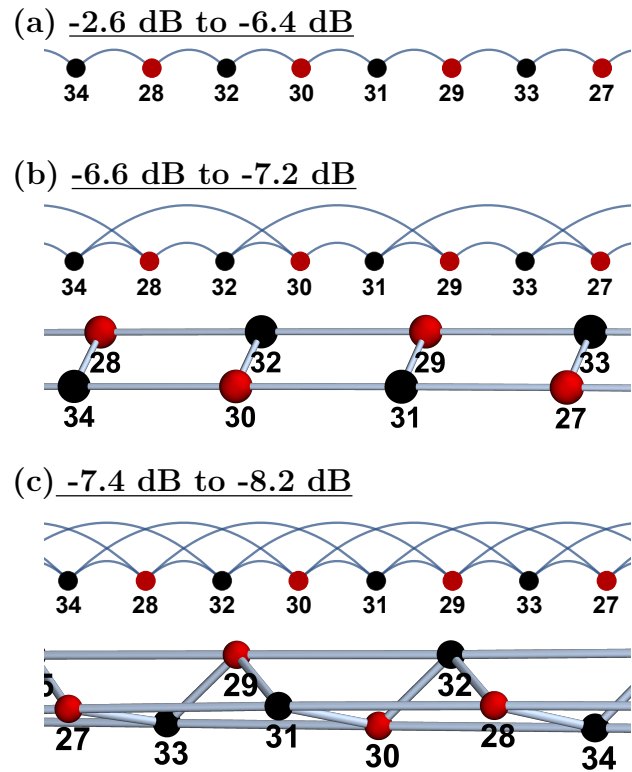 
It is remarkable that the lowest squeezing amounts already yield a 1D cluster wire spanning all qumodes in the considered set,  a universal structure for single qumode quantum processing. As the squeezing increases, the graphs gain ``width'' while retaining the 1D structure of the main two skew diagonals of the $A$ matrix, becoming a ladder structure then a spiraling wire overlapping with three parallel wires. Note the current record of optical squeezing is -15 dB~\cite{Vahlbruch2016}. Being a cluster state, the graph can be ``trimmed'' to a desired shape by measuring out the unwanted vertices~\cite{Briegel2001,Raussendorf2001}. Moreover, such structures has been shown to enable additional quantum processing options, also decided by measurements and feedforward within the model of one-way quantum computing~\cite{Alexander2014,Alexander2016a}. 

\section{Conclusion}

We have shown that a single OPO specifically engineered to add spatial degrees of freedom to the usual frequency ones of its resonant modes, can generate, in one fell swoop, sophisticated large-scale multipartite cluster states. We have derived analytic solutions for the state and have examined the physical significance of the edge weighting of the generated  graph state, in terms of the available experimental squeezing. Three essential points should be noted: 
\begin{itemize}
\item[{\em (i)}] These graphs being cluster states, they can be shaped and trimmed by measurements of undesired connected vertices.
\item[{\em (ii)}] Measurements and feedforward can also be used in a more elaborate manner, by taking advantage of the additional graph edge structure on a 1D or 2D backbone for implementing additional, arbitrary quantum operations~\cite{Alexander2014,Alexander2016a}.
\item[{\em (iii)}] Such cluster states can also be produced in parallel, by use of higher order spatial modes ${\lg _{ \pm 2,}}{\lg _{ \pm 3,}}...$ Although the effective coupling strength $\xi$ will initially be expected to be smaller for these modes due to the smaller overlap between the pump, signal and idler modes~\cite{Yang2016,Zhang2017},  $\xi$ can be effectively enhanced by introducing a specially designed nonlinear crystal structure for matching the property of LG modes and  the structured transverse mode of pump~\cite{Alves2018}.
\end{itemize}

\begin{acknowledgments}
This work was supported by National Key Research and Development Program of China(2017YFA0304502, 2016YFA0301404); National Natural Science Foundation of China (NSFC) (11874248,11874249); Natural Science Foundation of Shanxi Province (201801D121007). IK was supported by United States NSF grant DMR-1918207. CG-A, JH, and OP were supported by United States NSF grant PHY-1820882 and by the University of Virginia. The authors gratefully acknowledge Jacob Higgins and Prof.~Jiangrui Gao for stimulating discussions.  
\end{acknowledgments}

\appendix
\section{Analytical expressions of the A and G matrices}
Here we provide the details on how we obtain the eigenvalues/eigenvectors of the G matrix and the analytical expression of the A matrix. 
\subsection{Eigenvalues and eigenvectors of the G matrix}
Let us write the $2n\times2n$ G matrix, \eq G, as:
\begin{equation}
G = \left( {\begin{array}{*{20}{c}}
	0 & Q \\

    Q^{T} & 0 \\
	\end{array}} \right).
\end{equation}
	where:
\begin{equation}
Q=\text{  }\left(
\begin{array}{ccccc}
 \text{...} & 0 & 0 & 1 & 0 \\
 0 & 0 & \text{...} & 0 & 1 \\
 0 & 1 & \text{...} & \text{...} & 0 \\
 1 & 0 & 1 & 0 & 0 \\
 1 & 1 & 0 & 0 & \text{...} \\
\end{array}
\right)
\end{equation}
Let $M=QQ^T$. Note that $G^2$ has the form:
\begin{equation}
{G^2} = \left( 
\begin{array}{cc}
 Q Q^{T} & 0 \\
 0 & Q^{T} Q \\
\end{array}\right)
= \left( 
\begin{array}{cc}
M & 0 \\
 0 & J M J \\
\end{array}\right)
\end{equation}
 where $J$ is the anti-diagonal identity matrix,  $J_{ij}=\delta_{i,n-j+1}$. Here
 \begin{equation}
M=\left(
\begin{array}{cccccc}
 1 & 0 & 1 & 0 & \text{...} & 0 \\
 0 & 2 & 0 & 1 & 0 & \text{...} \\
 1 & 0 & 2 & 0 & \text{...} & 0 \\
 0 & 1 & 0 & \text{...} & 0 & 1 \\
 \text{...} & 0 & \text{...} & 0 & 2 & 1 \\
 0 & \text{...} & 0 & 1 & 1 & 2 \\
\end{array}
\right)
\end{equation}
It is possible to transform $M$ into a tridiagonal matrix $M'$ by a permutation of the indices, getting:
\begin{equation}
S M S^T\equiv M'=\left(
\begin{array}{cccccc}
 1 & 1 & 0 & 0 & 0 & \text{...} \\
 1 & 2 & 1 & 0 & 0 & 0 \\
 0 & 1 & 2 & \text{...} & 0 & 0 \\
 0 & 0 & \text{...} & \text{...} & 1 & 0 \\
 0 & 0 & 0 & 1 & 2 & 1 \\
 \text{...} & 0 & 0 & 0 & 1 & 2 \\
\end{array}
\right)
\end{equation}
where the permutation matrix S is defined by:
 \begin{equation}
(S^T\vec{\alpha})_j=\left\{\begin{array}{ll}
 
\alpha_i & j=2i-1\\
 
\alpha_{n-i+1} & j=2i
 
\end{array}\right.
\end{equation}
 As an example, for $n=12$, the $S$ matrix has the form:
 \begin{equation}
S = \left( {\begin{array}{*{20}{c}}
 	1 & 0 & 0 & 0 & 0 & 0\\
 	0 & 0 & 1 & 0 & 0 & 0\\
 	0 & 0 & 0 & 0 & 1 & 0\\
 	0 & 0 & 0 & 0 & 0 & 1\\
 	0 & 0 & 0 & 1 & 0 & 0\\
 	0 & 1 & 0 & 0 & 0 & 0\\
\end{array}} \right).
\end{equation}
The matrix $M^{\prime}$ is a simple tridiagonal matrix that has known eigenvalues and eigenvectors \cite{yueh2005eigenvalues}:
\begin{align}\label{eigsys M}
(v_k)_j&=\frac{2}{\sqrt{2n+1}}\sin\frac{k(2j-1)\pi}{2n+1}\\ %~~;~~
\lambda_k^2&=4\cos^2\frac{k\pi}{2n+1}.
\end{align}  
Since the eigenvalues of $M$ and $J M J$ are the same as of $M^{\prime}$'s, we conclude that the eigenvalues of $G^2$ are $\lambda_k^2$, with a double degeneracy, and for the matrix G the eigenvalues are 
\begin{equation}
\lambda_k=\pm2\left|\cos\frac{k\pi}{2n+1}\right|.
\end{equation}
 %, and the eigenvectors of $M$ are $S^T(v_k)_j$. 
We note that 
\begin{align}
[G,{\cal{J}}]&=0\\
\cal{J}&=\left(
\begin{array}{cc}
 0 & J \\
 J & 0 \\
\end{array}
\right)
\end{align}
and therefore the eigenvectors of $G$ can be chosen as eigenvectors of $\cal{J}$ as well, i.e. either symmetric or antisymmetric around the middle. We find that the symmetric eigenvectors are given by
\begin{align}\label{V sym} 
V_{k,\text{sym}}= \left(
\begin{array}{c}
 S^T v_k \\
 J S^T v_k \\
\end{array}
\right)\end{align}  with eigenvalues $(-1)^k {\lambda_k}$ and the antisymmetric ones are 
\begin{align}\label{V asym}  
V_{k,\text{asym}}=\left(
\begin{array}{c}
 S^T v_k \\
 -J S^T v_k \\
\end{array}
\right)\end{align}  with eigenvalue $(-1)^{k+1} {\lambda_k}$. 

\subsection{Construction of the A matrix.}
The $A$ matrix is obtained as follows~\cite{Menicucci2007}. %({\bf{need reference}}). 
First, diagonalize $G$ and separate into positive and negative blocks:
$$G=V D V^T,\text{     }D=\left(
\begin{array}{cccccc}
 \lambda _1 & 0 & 0 & 0 & \text{...} &
   \text{...} \\
 0 & \text{..} & 0 & 0 & \text{...} &
   \text{...} \\
 0 & 0 & \lambda _n & 0 & 0 & 0 \\
 0 & 0 & 0 & -\lambda _1 & 0 & 0 \\
 \text{...} & \text{...} & 0 & 0 &
   \text{..} & 0 \\
 \text{...} & \text{...} & 0 & 0 & 0 &
   -\lambda _n \\
\end{array}
\right)$$
Then:
$$
A=\left(
\begin{array}{cc}
 0 & A_0 \\
 A_0^T & 0 \\
\end{array}
\right)
$$
where
$$A_0=-V_{12}\left(V_{22}\right){}^{-1}\text{
      };\text{     }V=\left(
\begin{array}{cc}
 V_{11} & V_{12} \\
 V_{21} & V_{22} \\
\end{array}
\right)$$
Fortunately, in our case, $V$ has a simple form that allows the inversion of the relevant block.

To construct $V$ we use the eigenvectors \eqref{V sym} and \eqref{V asym} as column vectors, which will automatically yield a diagonalization of $G$. It is only left to order them so that the first columns correspond to positive eigenvalues and the second half of columns correspond to negative eigenvalues. To separate between positive and negative eigenvectors, we choose the ordering: $V_{1,asym},V_{2,sym},V_{3,asym},...$. 

Concretely, let 
\begin{align}
V_{11}=S^T( v_1,v_2,..,v_n)
\end{align}
Then our final form is of the form:
\begin{align}
V=\frac{1}{\sqrt{2}}\left(
\begin{array}{cc}
 V_{11} & V_{11} \\
 J V_{11} L & -J V_{11} L \\
\end{array}
\right)
\end{align}
where $L=diagonal(-1,1,-1,1,...)$ chooses the signs of the second half of columns in $V$ to correspond to the order of $asymmetric, symmetric, asymmetric, symmetric,...$ specified above.

The block $V_{22}=-J V_{11} L $ is clearly an orthogonal matrix, and we wind up with:
\begin{align}
A_0=-V_{12}\left(V_{22}\right)
   {}^{-1}=V_{11} L V_{11}^T J=S^TB^TSJ
\end{align}
where $B$ is built from the eigenvectors \eqref{eigsys M}:
\begin{align}
B_{ij}&=\frac{-4}{1+2n}\sum\limits_{k=1}^{n}(-1)^k\sin\frac{k(2i-1)\pi}{2n+1}\sin\frac{k(2j-1)\pi}{2n+1}\\
&=\frac{(-1)^{i+j+n}}{1+2 n}  \left[\frac{1}{\cos (\frac{(i-j) \pi }{1+2 n})}+\frac{1}{\cos (\frac{(i+j-1) \pi }{1+2 n})}\right],
\end{align}
where $i=1,2...,n;j=1,2,...,n.$
Finally, the $A$ matrix is eq A:
\begin{equation}
A = \left( {\begin{array}{*{20}{c}}
	0 & S^TB^TSJ \\
	JS^TBS & 0 \\
\end{array}} \right).
\end{equation}
\newpage
\bibliography{Pfister}
\end{document}